\documentclass[useAMS,usenatbib]{mn2e}
\usepackage{graphicx,epsfig,amssymb,amsmath,layout,verbatim,rotating,calc,mathrs
fs,natbib}
\usepackage{graphics}
\usepackage{rotate}
\usepackage{float}          
\usepackage{txfonts}

\newcommand{\ksk}{km~s$^{-1}$~kpc$^{-1}$ }    
\newcommand{\kms}{km~s$^{-1}$ }
\newcommand{\dgr}{$^{\circ}$}
\newcommand{\ej}{E$_{J}$ }
\newcommand{\x}{x1$^{\prime}$ }
\title[Slowly rotating bars]{ 
Morphologies introduced by bistability in barred-spiral galactic
potentials}
\author[L.~Tsigaridi et al.]
{L.~Tsigaridi,$^{1,2}$\thanks{liana@academyofathens.gr (LT);
patsis@academyofathens.gr (PAP)}
P.A.~Patsis$^{1}$\\
$^1$Research Center for Astronomy, Academy of Athens, Soranou Efessiou
    4, GR-115 27, Athens, Greece\\
$^2$ Section of Astrophysics, Astronomy and Mechanics, Department of
  Physics, University of Athens, Greece\\
}
\date{Accepted ..........Received .............;in original form ..........}

\begin{document}
\maketitle

\label{firstpage}
 
\begin{abstract}
We investigate the orbital dynamics of a \textit{barred-spiral} model when the
system is rotating slowly and corotation is located beyond the end of the spiral
arms.  In the characteristic of the central family of periodic orbits we find a
``bistable region''. In the response model we observe a ring surrounding the bar
and spiral arms starting tangential to the ring. This is a morphology resembling
barred-spiral systems with inner rings. However, the dynamics associated with
this structure in the case we study is different from that of a typical bar
ending close to corotation. The ring of our model is round, or rather elongated
perpendicular to the bar. It is associated with a folding (an ``S'' shaped
feature) of the characteristic of the central family, which is typical in
bistable bifurcations. Along the ``S'' part of the characteristic we have a
change in the orientation of the periodic orbits from a x1-type to a x2-type
morphology. The orbits populated in the response model change rather abruptly
their orientation when reaching the lowest energy of the ``S''. The spirals of
the model follow a standard ``precessing ellipses flow'' and the orbits building
them have energies beyond the ``S'' region. The bar is structured mainly by
sticky orbits from regions around the stability islands of the central family.
This leads to the appearance of X-features in the bars \textit{on} the galactic
plane. Such a bar morphology appears in the unsharp-masked images of some
moderately inclined galaxies. 
\end{abstract}

   
\section{Introduction}
\label{sec:intro}
In dynamical systems a ``bistability situation'' usually refers to cases where a
system has two stable equilibrium states. In a bifurcation diagram the curve of
steady state displays an ``S'' shape as a certain parameter of the system
varies. The ``S'' is delimited by two saddle-node bifurcation points. Between
them we have two stable and one unstable steady states \citep[see
e.g.][]{ang03,ly,strog14}. The corresponding situation in Hamiltonian Galactic
Dynamics is depicted in the characteristic of a family of periodic orbits as two
successive tangent bifurcations \citep[see e.g.][]{gco04} facing opposite
directions. These two bifurcations share the same unstable branch. In other
words the characteristic folds twice as the varying parameter, usually the
Jacobi constant \mbox{\ej\!}, increases. Foldings of the characteristic have
been encountered by \citet{spa02a} and \citet{spa02b}, in 3D Ferrers bar
potentials. However, the way they affect the face-on morphology of a model has
not been examined in those papers. Nevertheless, it was clear that the foldings
of the characteristics affect to a larger degree slowly rotating models. In this
paper we present the implications of the presence of such a folding of the
characteristic of the main family of periodic orbits for the dynamics of a
slowly rotating barred-spiral potential.

Slowly rotating models of disc galaxies have been proposed in the past to
describe the dynamics of normal (non-barred) spiral galaxies with open spiral
arms. In these models (stellar and gaseous) the symmetric, strong spiral
structure extends inside corotation \citep{cg86, cg88, pcg91, ksr03, metal04,
jetal13}. 

Contrarily, in barred galaxy models corotation is usually placed close to the
end of the bar \citep{gco80}. Recently \citet{fetal14} presented a list
with 32 barred galaxies in which they estimated the ratio of the corotation to
the bar radius, $R_c/R_b$, to be between $0.94\pm0.08 < R_c/R_b < 2.1\pm0.5$. 
Model bars ending well inside corotation have been found in N-body simulations
\citep{ce93} as well as in response models of barred potentials derived from
near-infrared observations \citep{rauetal08}. In all these studies, slowly
rotating bars are associated with late-type barred-spiral galaxies. It is
generally believed that bars in barred galaxies are supported by the x1 family
of elliptical, stable, periodic orbits, which extends along the major axis of
the bar \citep{cp80}, or, in the three-dimensional case, by the corresponding
families of the x1-tree, i.e. by x1 together with 3D families bifurcated at the
vertical resonances, \citep{spa02a}. The bar is built by trapping quasi-periodic
orbits along the orbits of these families. Deviations from this orbital
behaviour have been proposed, pointing to bars in which other families than x1
play a significant and perhaps a leading role in the building of the bars. Such
behaviours are favoured either in slowly rotating bar models in which the bars
end well before corotation \citep{pp86, spa02b}, or in bars with large major to
minor axis ratios \citep{kp05}. \citet{papo94} even claim, that in slowly
rotating bars of the type described by \citet{lb79}, there is a better matching
of the outer-to-inner ring radial ratio, than in standard fast rotating bars.
Finally, \citet{pkg10} have shown that bars of ansae-type can be supported
mainly by chaotic orbits in a (fast rotating) model based on the potential of
NGC~1300, estimated from near-infrared observations.

The implications of slow rotation for the orbital dynamics in a
\textit{barred-spiral} model, i.e. when the spiral component is explicitly
included in the potential, have not been extensively studied. Recently
\citet{tp13} have presented  a barred-spiral model, rotating with a single
pattern speed, characterized by a ratio $R_c/R_b$, about 2.9. This case, with
$\varOmega_p=15$~\ksk\!, has been considered as ``general'' since several
dynamical mechanisms cooperated in forming the obtained barred-spiral response
morphology. The action of two different dynamical mechanisms led to the
formation of an inner barred-spiral structure surrounded by an oval-shaped disc
and an outer set of spiral arms beyond corotation. However, if we decrease
further the pattern speed, there are even more considerable changes in the
orbital dynamics of the system. The pattern speed, $\varOmega_p$, is the most
important parameter for the resulting response morphology of the model.  

Besides the case presented in \citet{tp13}, we have studied a series of models
with even smaller $\varOmega_p$ down to 10~\ksk\!. As the pattern speed
decreases from $\varOmega_p=15$~\ksk\!, we have a rather new orbital behaviour,
which shapes a different barred-spiral response with distinct morphological
features. For example, despite the fact that the bars in all studied models have
comparable sizes, as $\varOmega_p$ decreases the orbital dynamics of the central
family of periodic orbits changes from that of the typical case \citep{cg89}. In
the present paper we present these changes, we describe the morphological
features that are encountered in extremely slowly rotating models, and we
discuss its relevance to some morphological features appearing in some images of
barred-spiral galaxies.

Our general model is the same as in \citet{tp13}. It is based on a modified
version of a potential estimated from the near-infrared photometry of the late
type barred-spiral galaxy NGC~3359 \citep{boony03, pkgb09}. We used it as a
general barred-spiral potential with $\varOmega_p$ being a free parameter. It is
worth noticing that \cite{ee85}, based on observation by \citet{sg82},
estimate the ratio of the semi-major axis of the bar to the length of the rising
part of the rotation curve for NGC~3359 to be $R_{bar}/R_{rise}=0.75$. According
to this work, the fact that the bar ends before the velocity curve turns over,
is an indication for slow bar rotation. Despite the difficulties in
estimating the inner parts of the rotation curves of barred galaxies, this
property is in general associated with slowly rotating bars, like the bars of
the models we study. In the series of our response models, since $R_{rise}
\approx 4.6$~kpc \citep[see figure 3 in][]{tp13} this ratio varies between 0.619
and 0.641. In the specific model of the present paper this ratio is about 0.62,
close to the lowest value. This means that our potential is adequate for
studying slowly rotating models in general.

The structure of the paper is the following: In Section~\ref{sec:model} we
briefly present again our potential \citep[for more details see][]{tp13}. The
results of our study are described in Section~\ref{sec:slow} and refer to the
building of the response features, which are the ring, the bar and the spirals.
These results are discussed in Section~\ref{sec:discuss} and 
in Section~\ref{sec:concl} we enumerate our conclusions.

\section{Summary of model properties}
\label{sec:model}
The model has been extensively described in \citet{tp13}. It is a two
dimensional
model of the general form
\begin{equation}
\label{eq:pot}
\Phi(r,\varphi) = \Phi_0(r) + \sum_{m=2,4,6} \Phi_{mc}(r)\,{\rm cos}\,(m\varphi) +
\Phi_{ms}(r)\,{\rm sin}\,(m\varphi)
\end{equation}
The components $\Phi_0$, $\Phi_{mc}$, and $\Phi_{ms}$ of the equation above are
given as polynomials of the form $\sum_n\!\!a_n r^n$, $n=0\dots8$. The radial
variation of the perturbation force normalized over the radial axisymmetric one
is
given in figure 1 in \citet{tp13}.

The equations of motion are derived from the Hamiltonian
\begin{equation}
\label{eq:hamilton}
H \equiv \frac{1}{2}\left(\dot{x}^{2} + \dot{y}^{2}\right) + \Phi(x,y) -
\frac{1}{2}\Omega_{p}^{2}(x^{2} + y^{2})=E_{J},
\end{equation}
where $(x,y)$ are the coordinates in a Cartesian frame of reference
rotating with angular velocity $\Omega_{p}$. $\Phi(x,y)$ is the
potential (\ref{eq:pot}) in Cartesian coordinates with the bar aligned
approximately with the y-axis, $E_{J}$ is the numerical
value of the Jacobi constant, hereafter called the energy, and dots denote time
derivatives.
\begin{figure}
\begin{center}
\resizebox{80mm} {!}{\includegraphics[angle=0]{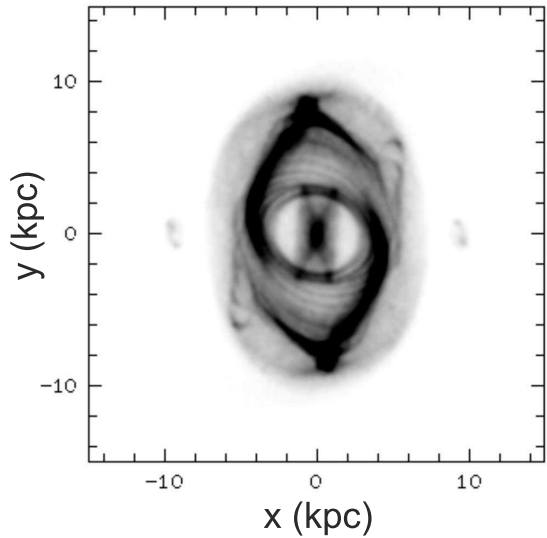}}
\end{center}
\caption {The stellar response model with $\varOmega_p = 11.5$~\ksk. The system
has completed 10 pattern rotations, rotating counter-clockwise. The bar
extends roughly along the y-axis. Corotation is at $R_c \approx$ 10~kpc} 
\label{resp11} 
\end{figure}
\begin{figure}
\begin{center}
\resizebox{84mm} {!}{\includegraphics[angle=0]{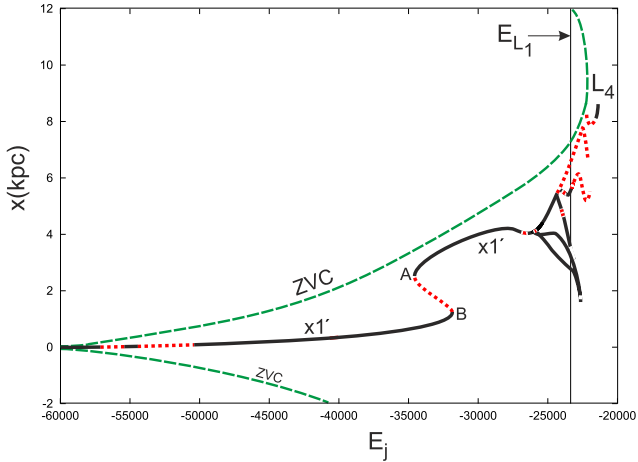}}
\end{center}
\caption {The characteristics of \x and its bifurcations for $\varOmega_p =
11.5$~\ksk\!\!. The dashed line, labelled ``ZVC'', is the curve of zero
velocity. Continuous black parts of the characteristic indicate stability, while
dotted parts, red in the online version, instability. The $E_{L_1}$ line is
indicated with an arrow in the upper right corner of the figure.}
\label{xar1a} 
\end{figure}

\section{Slowly rotating models}
\label{sec:slow}
By varying $\varOmega_p$ between $10 < \varOmega_p < 30$~\ksk\!, we obtained
always a kind of barred-spiral response. In this range of pattern speeds
the corotation radius of the models, $R_c$, takes values between $12 \gtrsim R_c
\gtrsim 4.3$~kpc  respectively.  Nevertheless, while the pitch angle of the
response spirals varied considerably in models with different pattern speeds (it
was larger for lower pattern speeds), the radius of the response bar varied only
between $2.85 < R_b < 2.95$~kpc. For $\varOmega_p > 30$~\ksk the size of the
response bar was clearly decreasing. For example, for $\varOmega_p = 35$~\ksk we
estimated it to be about 2.45~kpc. 

The changes that are introduced in the dynamics of the system as $\varOmega_p$
decreases are reflected in changes observed in the (\ej\!\!,$x)$ characteristic
curve \citep[for a definition see][section 2.4.3]{gco04} of the central family.
In typical cases of barred galaxy models, the $x_0$ initial condition of the
central, x1, family is increasing with \ej between the inner Lindblad Resonance
(ILR) and the 4:1 resonance \citep{cg86, cg89}. This is the case also for the
present model for $\varOmega_p > 16$~\ksk$\!\!.$ However, for slower rotating
models, i.e. for $\varOmega_p \lessapprox 15$~\ksk~\!\!\!, we observe a folding
of the characteristic curve well before corotation in the (\ej\!\!,$x)$ diagram.
The $\Delta$\ej range over which we have the folding in a model increases with
deceasing $\varOmega_p$. We call this feature, the ``S''. As we will see,
foldings of the characteristics introduce in the system new orbital dynamics
accompanied by new morphological features. For this reason we call just in this
work the central family of our model x1$^{\prime}$ in order to distinguish it
from the standard x1 family, which has as members only elliptical periodic
orbits elongated along the bar. 

In the present paper we describe these changes in a typical case with
$\varOmega_p = 11.5$~\ksk\!\! (chosen so that we have 
$R_c \approx$ 10~kpc). For this $\varOmega_p$ the morphological features
associated
with slow rotation dominate and this facilitates the description of the relevant
dynamical mechanisms. 
 
The stellar response model for the potential (\ref{eq:pot}) with $\varOmega_p =
11.5$~\ksk is given in Fig.~\ref{resp11}.  The response bar length is $R_b
\approx 2.85$~kpc.  The system has completed 10 pattern rotations. Initially the
particles have been distributed randomly within a disk of radius
r$_{max}$=11~kpc with velocities securing circular motion in the axisymmetric
part of the potential, $\Phi_0(r)$. At the beginning of the simulation the
amplitudes $\Phi_{mc}$ and $\Phi_{ms}$ of the perturbing term grow linearly
with time from 0 to their maximum value within two system periods (``time
dependent phase'') and after that
they remain constant. The time-dependency of the amplitudes in the beginning of
the simulation secures a smooth response of the system, since the initial
velocities are for circular motion in the axisymmetric term $\Phi_0(r)$.

\begin{figure}
\begin{center}
\resizebox{80mm} {!}{\includegraphics[angle=0]{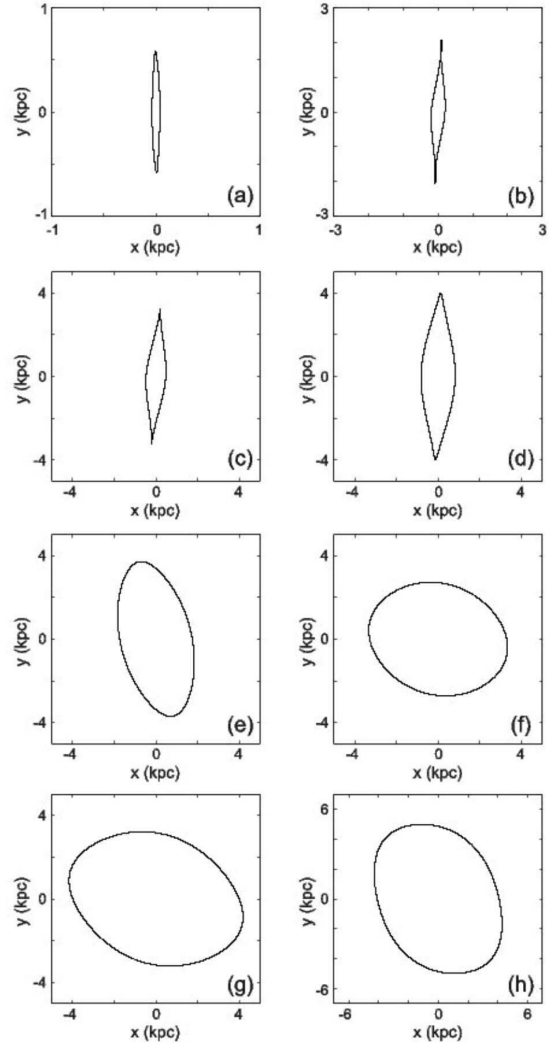}}
\end{center}
\caption {Periodic orbits along the characteristic of \x for \ej = $-55000$ (a),
$-45000$ (b), $-37000$ (c), $-33000$ (d),(e),(f) (the three \x representatives
at the ``S'' region), $-30000$ (g) and $-27000$ (h). Note the different scales
of the axes.}
\label{orbchar} 
\end{figure}
The snapshot has been converted to an image using the ESO-Midas software. In the
central part we observe a rectangular shaped bar. The contrast of the image has
been chosen such as to allow us to clearly see that inside the rectangular
shaped bar appears an ``X'' feature. The bar, with the embedded in it X-feature,
ends on an oval (pseudo)ring structure. This ring has a certain width. It is
almost round at its inner boundary, which is attached to the bar, while the
ellipticity of the orbits that form it increases outwards and becomes maximum at
its outer part, which coincides with the beginning of the spiral structure. At
this point the ring is elongated in a direction roughly vertical to the bar (at
an angle about $15^{\circ}$ with respect to the x-axis).  The length of the
semimajor axis of this ring is about 3.6~kpc. On the sides of the bar, inside
the ring, the response surface density is very low. The double armed open spiral
structure starts tangentially from the ring  having a pitch angle $i_p \approx$
35\dgr in the outer parts (it is not logarithmic). The spiral arms have a sharp
bifurcation at $r \approx 7.8$~kpc.
\begin{figure*}
\begin{center}
\resizebox{160mm} {!}{\includegraphics[angle=0]{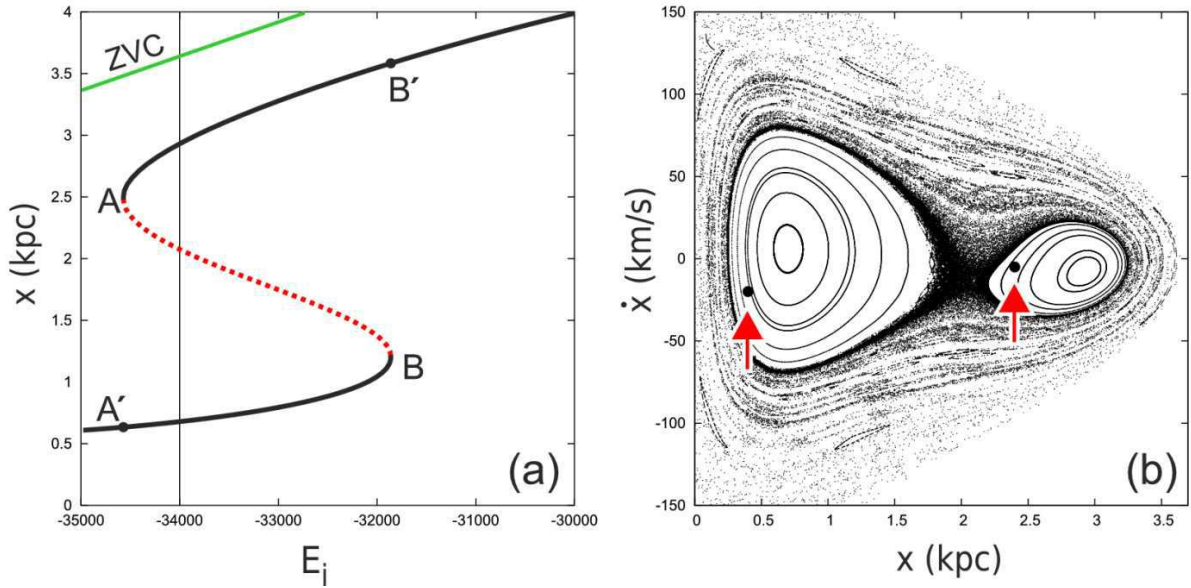}}
\end{center}
\caption{The evolution of the surfaces of section in the "S" region. In (a) we
focus in the "S" region of the characteristic. Continuous black parts of the
curve indicate stability, while dotted (red in the online version) indicate
instability. The vertical line at \ej = $-34000$ denotes the energy at which we
give the $(x,\dot{x})$ surface of section in (b). With increasing energy between
A, A$^{\prime}$ and B, B$^{\prime}$ in (a) the size of the left stability island
decreases, while the size of the right one increases. The arrows point to the
orbits depicted in Fig.~\ref{quasi34}.} 
\label{ssos34} 
\end{figure*}

The (\ej\!\!,x) characteristic curve of \x and its bifurcations, as we approach
L$_4$, is given in Fig.~\ref{xar1a}. Continuous black parts of the curves
indicate stability, while dotted (red in the online version) indicate
instability. The stable Lagrangian points L$_4$ and L$_5$ are at energy
E$_{L_{4,5}}=-22196$ (\ej is always given in units of km$^{2}$ s$^{-2}$). L$_4$
in Fig.~\ref{xar1a} is at the local maximum of the curve of zero velocity at the
right side of the figure. For the unstable Lagrangian points L$_1$ and L$_2$ we
have E$_{L_{1,2}}=-23360$. At this energy we have drawn a solid vertical line in
Fig.~\ref{xar1a} and we indicate it with an arrow labelled ``E$_{\rm{L_1}}$''.
The folding of the characteristic curve, which we call ``S'', occurs between
$-34500 < $ \ej $< -31800$ (\ej $\approx -34500$ is the energy at point ``A''
and \ej $\approx -31800$ is the energy at point ``B''). 

Since there is no gap or discontinuity in the curve, we consider the whole curve
belonging to a single family of periodic orbits, namely the \x. The \x family is
stable close to the centre of the model, then it becomes unstable for $ -57000
<$ \ej $< -50436$ (with a small stability interval $ -55000 <$ \ej $< -54336$).
After that it remains practically stable until the region of the ``S''. For \ej
$> -31800$ (\ej at point ``B''), which we consider as the end of ``S'', the
upper branch of the characteristic remains stable until the energy \ej $\approx
-27000$. Beyond this energy follows a tree of bifurcating families. 

Up to this point, the morphological evolution of the periodic orbits along the
\x characteristic curve shows an interesting variation. In Fig.~\ref{orbchar} we
give successively characteristic periodic orbits as the energy increases.
\begin{figure*}
\begin{center}
\resizebox{150mm} {!}{\includegraphics[angle=0]{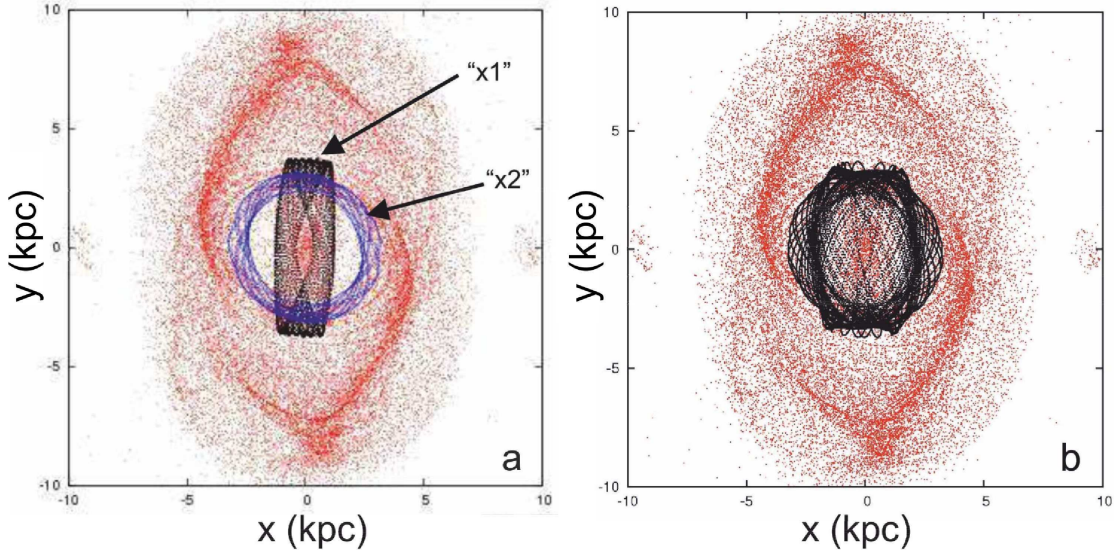}}
\end{center}
\caption{(a) Two characteristic quasi-periodic orbits from the "S" region of \x
supporting
a x1 and a x2 flow. Both are at \ej=$-34000$. They are overplotted on our
response model. The location of their initial conditions on the $(x,\dot{x})$
surface of section is indicated in Fig.~\ref{ssos34}b. (b) Three orbits from
the chaotic region between the stability islands of Fig.~\ref{ssos34}b plotted
in the central region of the model. Instead of a ring they contribute to the
formation of a bulge-like central component.}
\label{quasi34} 
\end{figure*}
For \ej $\lessapprox -54000$ the \x orbits are typical x1 ellipses, elongated
along the y-axis, as in Fig.~\ref{orbchar}a. Then for $-54000\lessapprox$ \ej
$\lessapprox -38000$ they develop loops at their apocentra
(Fig.~\ref{orbchar}b). As energy increases further, for \ej $> -38000$, the
loops at the apocentra vanish again (Fig.~\ref{orbchar}c) and the morphology of
the periodic orbits as we approach the ``S'' region is as in
Fig.~\ref{orbchar}d. In the ``S'' region we have three simple periodic orbits at
each energy. Two stable and one unstable. At the local maximum of ``S'' (\ej
$\approx -31800$), where the characteristic turns to the left (point ``B'' in
Fig.~\ref{xar1a}), the orientation of the elliptical periodic orbits starts
changing, while, simultaneously, they become unstable. Their major axes lean
more and more towards the x-axis as energy decreases. For example the orbit in
Fig.~\ref{orbchar}f, at \ej $\approx -34500$ (point ``A ''), the major axis of
the periodic orbits is close to the x-axis (minor axis) of the system. In other
words, moving along the unstable segment of the characteristic from ``B'' to
``A'' we change from a x1- to a x2-like \citep[see e.g.][]{cg89} orientation.
Then, moving again from \ej $\approx -34500$ to the right along the upper stable
branch of ``S'', the orbits become stable and of ``x2-type'' until \ej $\approx
-30000$ (Fig.~\ref{orbchar}g). Beyond this point we have periodic orbits like
the one in Fig.~\ref{orbchar}h. If we plot together successive \x periodic
orbits with increasing \ej for \ej $> -28000$ we observe that their major axes
start tilting towards the y-axis this time, building a ``precessing ellipses
pattern'' \citep{p08} that can be considered as the backbone of a spiral
structure extending to larger distances. The spiral structure is discussed below
in section~\ref{sec:spires}.

\subsection{The ring}
\label{sec:ring}
In Fig.~\ref{ssos34} we focus into the ``S'' region. The corresponding part of
the characteristic is given in Fig.~\ref{ssos34}a and is included between the
energies of ``A'', ``A$^{\prime}$'' and ``B'', ``B$^{\prime}$'. ``A'',
``A$^{\prime}$'' are at \ej$\approx -34500$ and  ``B'', ``B$^{\prime}$'' are at
\ej$\approx -31800$. At any energy between, e.g. for \ej $=-34000$, where we
have drawn a vertical line in Fig.~\ref{ssos34}a, we have three periodic orbits.
There is always an unstable \x periodic orbit with an intermediate inclination
between the stable x1-like (lower part of the characteristic) and the stable
x2-like (upper part of the characteristic). The situation can be described as a
case of a bistable bifurcation. We can say that at points ``A'' and ``B'' we
have a direct and an inverse tangent bifurcation.

\begin{figure}
\begin{center}
\resizebox{80mm} {!}{\includegraphics[angle=0]{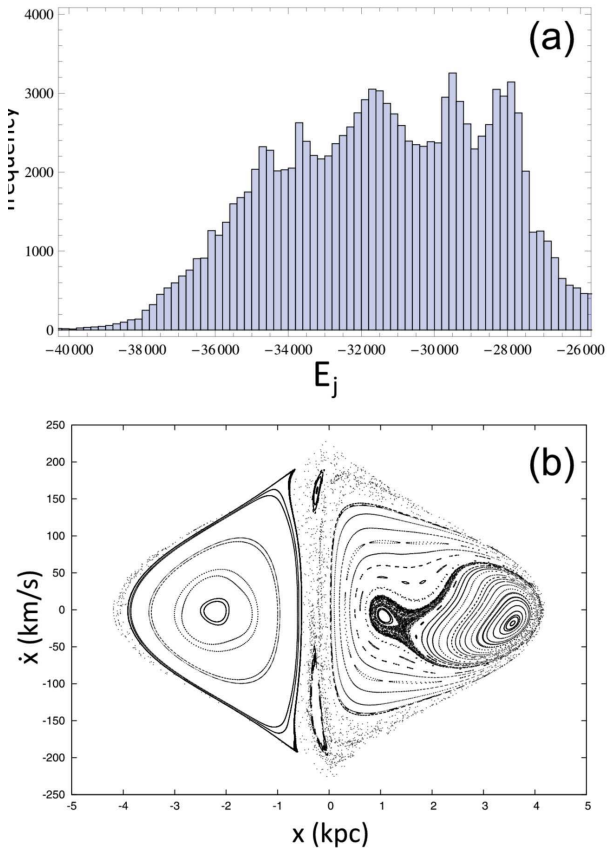}}
\end{center}
\caption{(a) A histogram showing the \ej distribution of the particles with
$2.5< r < 3.8$~kpc in Fig.~\ref{resp11}. (b) The $(x,\dot{x})$ surface of
section at \ej=$-32000$, which is a local maximum in (a).}
\label{histring} 
\end{figure}
In Fig.~\ref{ssos34}b we give the $(x,\dot{x})$ surface of section for the
energy \ej $=-34000$. It is a typical case of a surface of section in the ``S''
region. At the centre of the left stability island we have the ``x1'' periodic
orbit, while at the centre of the right stability island we have the x2-like
ellipse. The extent of this cross section, as well as of all other similar
figures we discuss in our paper, is limited by the zero velocity curve, ZVC
(drawn in Fig.~\ref{ssos34}a). Moving from ``B'' to ``A'' along the unstable
part of the characteristic in Fig.~\ref{ssos34}a the area of the right stability
island in the surface of section (Fig.~\ref{ssos34}b) decreases, while the area
of the left one increases (Fig.~\ref{ssos34}b is close to the left border of the
``S''). At \ej $\approx - 33000$ the width of the chaotic zone between the two
ordered regions becomes minimum. The role of the x2-like orbits is even more
emphasised as \ej increases  beyond ``B$^{\prime}$'' along the characteristic in
Fig.~\ref{ssos34}b. From ``B$^{\prime}$'' until \ej $\approx -30000$, the only
simple periodic orbits existing are almost perpendicular to the bar. Since they
are stable they may attract around them quasi-periodic orbits. If these orbits
are populated, they will support a x2 flow. 

By inspection of Fig.~\ref{orbchar} we can understand the association between
the \x periodic orbits and the non-periodic orbits that have been populated in
the energy interval of the ``S'' region (Fig.~\ref{ssos34}) in order to give the
response morphology depicted in Fig.~\ref{resp11}. We remind that the initial
conditions of the particles in the response model are distributed randomly on a
disc of radius 11~kpc and take velocities for circular motion in $\Phi_0$ in
Eq.~(1). The orbits that populate the response model have as initial conditions
the positions and the velocities of the test particles at the end of the ``time
dependent phase'' of the model that lasts two rotational periods.
In general the \x periodic orbits with small initial $x_0$ values ($0.6
\lessapprox x_0 \lessapprox 1.1$) are orientated along the bar
(Fig.~\ref{orbchar}a-d). They are ellipses with their major axes roughly aligned
with the major axis of the bar. However, the projections of such orbits from the
``S'' region, as well as the projections of the quasi-periodic orbits trapped
around them, on the major axis of the bar (y-axis) are larger with respect to
the bar of the response model, which is included inside the ring
(Fig.~\ref{resp11}). If we consider quasi-periodic orbits from the left
stability island of the surfaces of section in the ``S'' region they always
exceed the size of the ring of the response model. Contrarily, the size of the
quasi-periodic orbits from the right stability island is such as to contribute
to the formation of the ring, without exceeding the dimensions of the response
feature for all \ej values along the ``S''. Two quasi-periodic orbits for \ej
$=-34000$ that illustrate the above statement are given in Fig.~\ref{quasi34}a.
The locations of the orbits drawn in Fig.~\ref{quasi34}a in the corresponding
stability islands are indicated with arrows in Fig.~\ref{ssos34}b. The one with
the small $x_0$ in Fig.~\ref{ssos34}b is indicated with ``x1'' in
Fig.~\ref{quasi34}a.

\begin{figure}
\begin{center}
\resizebox{80mm} {!}{\includegraphics[angle=0]{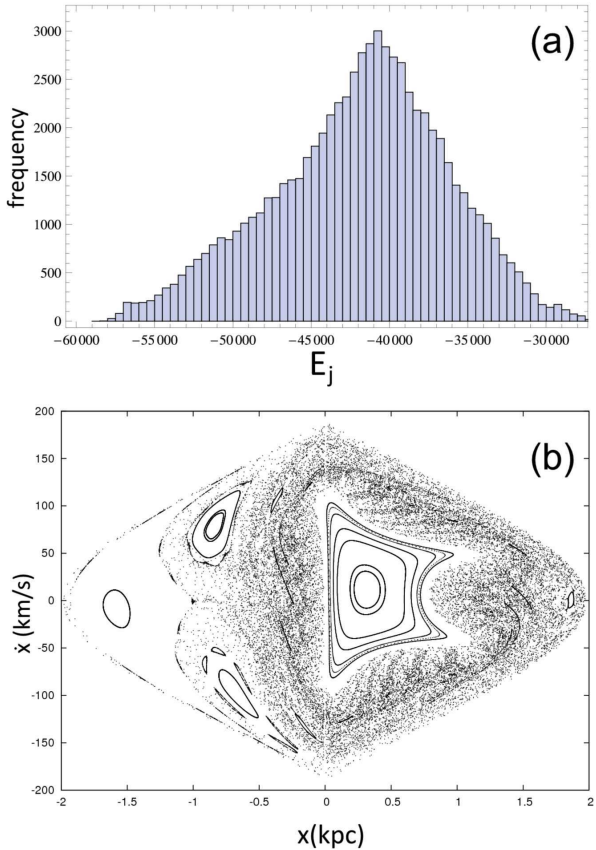}}
\end{center}
\caption{In (a) we give the \ej distribution of particles located at $r <
2.5$~kpc in our response model (Fig.~\ref{resp11}. The mode is at \ej $=
-40750$. In (b) we give the $(x,\dot{x})$ surface of section at the energy of
the mode.}
\label{mode} 
\end{figure}
Chaotic orbits associated with the unstable branch of the ``S'' are  also not
populated in the response model (Fig.~\ref{resp11}). In the case depicted in
Fig.~\ref{ssos34}b, such orbits are found in the chaotic zone between the two
stability islands. Integrated for 8 rotational periods, they fill the whole
central region of the model, including the rather empty areas between the bar
and the beginning of the spiral arms, inside the ring. The presence of these
orbits in the response model would build a kind of response bulge, instead of a
ring. This can be seen in Fig.~\ref{quasi34}b, where we plot on top of the
model, three orbits from the chaotic region with initial conditions $(x,\dot{x})
= (2.1,-5)$, $(1.7,-35)$ and $(1.9,-35)$. We observe that they fill the interior
of the ring, unlike with what happens in the model. So we can exclude them
from the orbits contributing to the response morphology observed in 
Fig.~\ref{resp11}.

The identification of the ring with the x2-like orbits in the ``S'' region
(upper stable branch of ``S'') can be verified by considering the distribution
of the energies of the particles located in the annulus with $2.5 < r < 3.8$~kpc
in the response model. This is roughly the region of the ring. The radius
$r=2.5$ is reaching the inner part of the drawn ``x2'' orbit in
Fig.~\ref{quasi34}a, while $r=3.8$ reaches the drawn apocentra of the ``x1''
orbit and the beginning of the spiral arms in the same figure. This energy
distribution is given in Fig.~\ref{histring}a. We observe that most of the
particles in the ring have energies in the range $-35000\lessapprox$ \ej
$\lessapprox -28000$. Particles with \ej $> -28000$ are in quasi-periodic orbits
trapped around the stable periodic orbits participating in the beginning of the
``precessing ellipses'' pattern that builds the spirals (see also
Section~\ref{sec:spires} below). Their small contribution in the distribution of
the energies in the annulus can be seen at the right part of
Fig.~\ref{histring}a for \mbox{\ej $> -28000$}. For \ej $> -32000$ we do not
have \x orbits with orientation along the major axis of the bar. This means that
if we find in the ring area  particles in bar-supporting x1-like orbits they
should have \ej$<-32000$. However, as we have seen, we have excluded the
presence of x1-like orbits from the lower branch of ``S'' in the response model,
since their size exceeds the size of the ring. So, according to
Fig.~\ref{histring}a, if such orbits exist, they will be a minority with
\ej$<-34500$. We also note that as we move
in the lower branch of the ``S'' from ``A$^{\prime}$'' to B (Fig.~\ref{ssos34}a)
the importance of the orbits following the x1 flow decreases. In
Fig.~\ref{histring}b we can see the relative importance of the island around the
x2-like stable orbit, centred at $(x,\dot{x}) = (3.56,-0.1823)$ with respect to
the island around the x1-like, centred at $(x,\dot{x}) = (1.05,-8.2)$, at \ej $=
-32000$. The local maximum of the histogram at about \ej $\approx -32000$
reflects the increasing importance of the stability island around the x2-like
stable orbits as we approach this \ej value. 

This analysis shows that the observed ring structure in our response model is
due to the x2-like orbits that are introduced in the system in the energy range
where we have the ``S'' feature in the characteristic in the bistable region of
the central family of periodic orbits. Stable periodic orbits with x2-like
orientation exist on the \x characteristic also beyond ``B$^{\prime}$'' in
Fig.~\ref{ssos34}a (up to \ej$\lessapprox -30000$ and provide to the system the
backbone of the ring. The presence of the ``S'' feature in the characteristic is
also associated with the end of the contribution of orbits trapped around stable
x1-like orbits aligned with the y-axis to the response morphology of the model.
As we have seen in Fig.~\ref{histring}a their contribution to the ring region is
minor. Thus, the bar of the model ends practically at the ring. In general up to
the energy of ``A'' and ``A$^{\prime}$'' (\ej $\approx -34500$) our response
model is populated by orbits associated with the lower branch of the \x
characteristic. Then we have a jump from ``A$^{\prime}$'' to ``A'' and the
backbone of our model are the stable orbits along the upper branch of the
characteristic.
\begin{figure*}
\begin{center}
\resizebox{120mm}{!}{\includegraphics[angle=0]{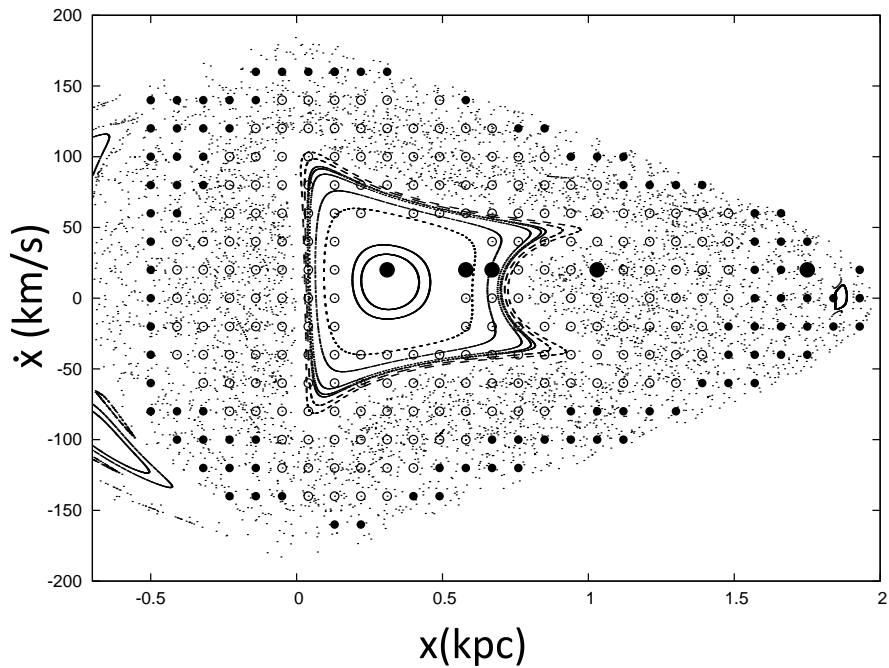}}
\end{center}
\caption{The initial conditions of the integrated orbits on the $(x,\dot{x})$
surface of section at \ej=$-40750$. Small empty bullets correspond to orbits
supporting a narrow X feature matching the dimensions of the response bar.
Small filled bullets correspond to orbits with an overall X morphology but not
matching the dimensions of the bar. The large black bullets indicate the initial
conditions of the orbits depicted in Fig.~\ref{modeorbs}.}
\label{sosmode} 
\end{figure*}
\subsection{The bar}
\label{sec:bar}
Let us focus now in the response bar and its internal structure. If we confine
ourselves to the study of the region of the response model
(Fig.~\ref{resp11}) with $r < 2.5$~kpc, we practically select the region of the
bar. The majority of
the particles with $r < 2.5$~kpc are located in the bar, since the regions to
its sides, inside the ring, are regions almost depleted of particles. The energy
distribution in this region is given in Fig.~\ref{mode}a. We observe that
practically we have particles with \ej $< -30000$ and that the contribution of
particles with \ej $> -35000$, i.e. to the right of A, A$^{\prime}$ in
Fig.~\ref{ssos34}a, is small. The mode of the distribution is at \ej $= -40750$.

The main feature of the response bar is an ``X'' feature, discernible in
Fig.~\ref{resp11}. We try to understand the mechanism leading to its formation
first by investigating the possible contribution of particles with the energy of
the mode in Fig.~\ref{mode}a.  The $(x,\dot{x})$ surface of section at this
energy is given in Fig.~\ref{mode}b. The stability island that dominates in the
$x>0$ region is the island of \x\!, which in this energy is x1-like. It is
surrounded by a chaotic sea that extends also to $x<0$. At the left side of this
chaotic sea we observe two more stability islands belonging to two families of
periodic orbits that have a 3:1 character and their characteristic joins the
branches of the \x characteristic at energies beyond the ``S'' in
Fig.~\ref{xar1a}. These orbits do not play any particular role in our study, so
we will not deal further with them.

\begin{figure*}
\begin{center}
\resizebox{160mm} {!}{\includegraphics[angle=0]{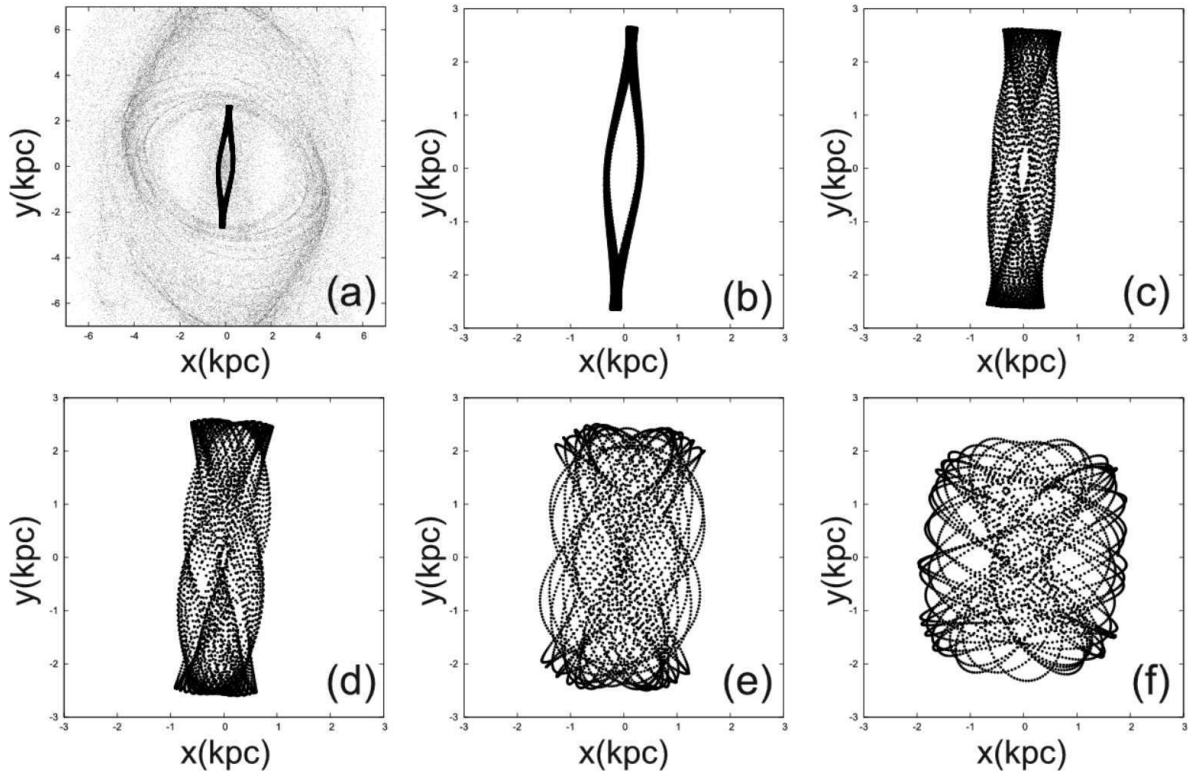}}
\end{center}
\caption{A series of non-periodic orbits with \ej=$-40750$, integrated for
10 pattern periods, that demonstrate the orbital morphologies encountered at
this energy. Their initial conditions are indicated with large bullets in
Fig.~\ref{sosmode}. All of them have $\dot{x_0}$=20~\kms\!\!, while from (a) to
(f)
their $x_0$ are 0.31, 0.31, 0.58, 0.67, 1.03 and 1.75 respectively. The scale
of the orbits is the same for panels (b) to (f).}
\label{modeorbs} 
\end{figure*}

In order to find out the orbits that support the ``X'' we consider a grid of
initial conditions with a step 0.09 in the x- and 20 in the y- direction on the
surface of section of Fig.~\ref{mode}b. We integrated the orbits with initial
conditions on the nodes of the grid for 30 pattern periods. The initial
conditions of the integrated orbits that have been found to give orbits with
morphologies relevant to our study are given with filled and empty bullets in
Fig.~\ref{sosmode}. They are encountered for $x> 0.5$. In order to demonstrate
their morphologies, we present below some characteristic orbital shapes. The
large black bullets in Fig.~\ref{sosmode} give the initial conditions of five
typical orbits that are described in Fig.~\ref{modeorbs}. All of them have
$\dot{x_0}$=20~\kms\!\!, while their $x_0$ initial condition increases from
Fig.~\ref{modeorbs}a to Fig.~\ref{modeorbs}f. The large bullet inside the
innermost invariant curve in Fig.~\ref{sosmode} ($x_0$=0.31) is given in
Fig.~\ref{modeorbs}a and Fig.~\ref{modeorbs}b. In Fig.~\ref{modeorbs}a it is
drawn on top of the response model, so that we can see its extent relative to
the bar and the ring structure. Its morphology can be seen in another scale in
Fig.~\ref{modeorbs}b and is similar to the morphology of a x1-like periodic
orbit. For $x_0$=0.58, we are still in the area of the stability island
(Fig.~\ref{modeorbs}c). The orbit has a boxy character, while we observe the
appearance of wings of a X feature being formed towards the apocentra of the
orbit. On the central stability island of Fig.~\ref{sosmode}, but close to its
last KAM curve, the hole of the orbit around (0,0) almost vanishes and we have a
fully developed X feature in the interior of the orbit. This can be seen in
Fig.~\ref{modeorbs}d, with $x_0$=0.67. Surrounding the stability island we find
a plethora of chaotic orbits, that integrated at least for 10 pattern periods
reinforce the appearance of the X. In most cases the X feature was clearly
discernible even after integrating the orbit for 30 pattern periods. A typical
case is the one given in Fig.~\ref{modeorbs}e. For this chaotic orbit we have
$x_0$=1.03. The dimensions of this and all other similar chaotic orbits
correspond to the dimensions of the response bar we see in Fig.~\ref{resp11}.
Finally, we find that starting integrating chaotic orbits at the outer border of
the chaotic sea, close to the curve of zero velocity, we find again orbits with
an X feature embedded in them. However, these orbits are rounder than all other
orbits we discussed up to now. A characteristic example can be seen in
Fig.~\ref{modeorbs}f ($x_0$=1.75). Due to their stubby morphology they do not
contribute to the building of the structure of the response bar, since they
extend further to its sides.

Using an algorithm similar to that described in \citet{chpb11}, we applied
simple criteria to characterize the orbits according to their morphology and the
degree they support the X-shaped bar of the response model in Fig.~\ref{resp11}.
In Fig.~\ref{sosmode} small bullets indicate only the orbits we find supporting
an X feature in the bar region. The empty small bullets indicate narrow orbits
that remain confined in the region of the bar ($|x|\lessapprox 1.5$). Such
orbits are like those depicted in Fig.~\ref{modeorbs}c,d and e. With filled
small bullets we indicate orbits similar to the one in Fig.~\ref{modeorbs}f.
They have an interior X-shaped structure, though not as sharp as in
Fig.~\ref{modeorbs}e, but their dimensions do not match the dimensions of the
bar. In conclusion the X-shaped bar is supported mainly by particles in orbits
with initial conditions corresponding to the open bullets in
Fig.~\ref{sosmode}. 

By inspection of Figs.~\ref{mode},\ref{sosmode} and \ref{modeorbs} we first
realize that the initial conditions supporting the response bar morphology
within a certain time are distributed on the surface of section independently of
the locations of stability islands or chaotic seas. In general the stability
region on Poincar\'{e} cross sections that is excluded from participating in the
building of a bar is found around the retrograde family x4 \citep[e.g.][]{cg89}.
However, in studies of barred-spiral potentials estimated from near-infrared
observations of galaxies \citep{chpb11,tp13}, it has been realized that there
are initial conditions on x1 stability islands, which do not support
a particular morphological feature of a bar. On the other hand, we have found
that there are initial conditions in the chaotic seas that reinforce a
particular bar morphology \citep{paq97,pkg10,chpb11,tp13}. In the present case
quasi-periodic orbits that support part of the wings of an X feature are
encountered at the periphery of the stability island of the x1-like periodic
orbit. As noticed in \citet{p05} the morphology of the periodic orbit at the
center of an island may differ from the morphology of the quasi-periodic orbits
of the outer invariant curves. For example, in cases of elliptical x1-like
periodic orbits, the quasi-periodic orbits at the edge of the stability island
may support an ansae-type morphology. As we see in
Fig.~\ref{modeorbs} the morphology of the innermost quasi-periodic orbits, which
is similar to that of the periodic orbit itself, does not resemble that of an X.
On the other hand the integrated for 10-30 pattern periods chaotic orbits with
initial conditions in the chaotic sea surrounding the stability island have a
morphology matching that of the response bar (cf. Fig.~\ref{resp11} and
Fig.~\ref{modeorbs}e). 
\begin{figure}
\begin{center}
\resizebox{70mm} {!}{\includegraphics[angle=0]{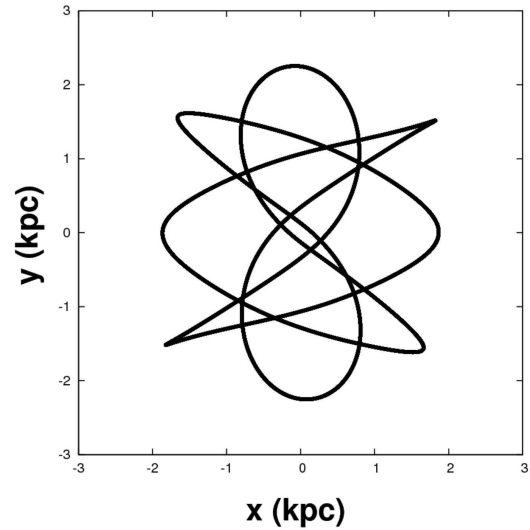}}
\end{center}
\caption {The periodic orbit of multiplicity 3 that has around it a sticky
zone with chaotic orbits that have for several tens of pattern periods a
morphology similar to the orbit in Fig.~\ref{modeorbs}f.} 
\label{tritone} 
\end{figure}
\begin{figure}
\begin{center}
\resizebox{80mm} {!}{\includegraphics[angle=0]{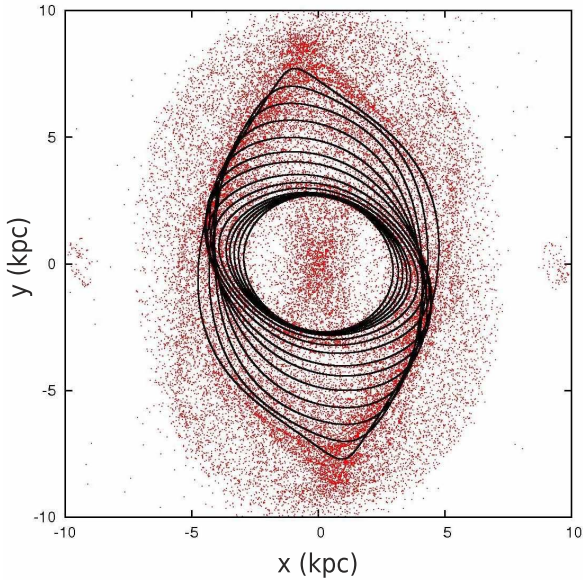}}
\end{center}
\caption{Periodic orbits overplotted on the response model in the
range $-34000 < $\ej $< -25000$. These orbits are the
backbone for the support of the observed spiral pattern and the ring pattern.
Energies from inside to outside:
$-34000$,$-33500$,$-33000$,$-32500$,$-32000$,$-31000$,$-30000$,$-29000$,
$-28000$, $-27500$, $-27000$, $-26500$,$-26000$,$-25500$ and $-25000$.}
\label{rsporbs} 
\end{figure}
Since morphologies similar to that of Fig.~\ref{modeorbs}e are encountered by
integrating the initial conditions corresponding to the open bullets in
Fig.~\ref{sosmode}, we can compare the area these open bullets occupy with the
density of the consequents on the surface of section in Fig.~\ref{mode}b. It
becomes apparent that the open bullets symbols are located in the sticky
region, i.e. the region with larger density,
surrounding the central stability island in Fig.~\ref{mode}b. As regards
the study of the orbital morphology it is worth to underline that moving from
the center of the stability island outwards and then crossing the surrounding
sticky region, we have a smooth morphological transition from narrow
non-periodic orbits, along the x-direction, to broader non-periodic orbits,
which harbour an X feature. The energy \ej=$-40750$ is the mode of the
distribution in Fig.~\ref{mode}a. However, a similar analysis can be done for
every \ej roughly in the interval $-45000 < $\ej $< -35000$ (practically for
energies before the ``S'' feature). The non-periodic orbits supporting the X are
mainly orbits in the sticky zones surrounding stability islands of x1-like
periodic orbits.

The zone with the small filled bullets in Fig.~\ref{sosmode} is a sticky zone as
well, this time around a periodic orbit of multiplicity 3. Its three tiny
stability islands can be better seen in Fig.~\ref{mode}b at $(x,\dot{x}) \approx
(1.8,0)$, $(-0.2,-130)$ and $(-0.4,110)$. In Fig.~\ref{tritone} we give this
periodic orbit, so that its morphological relationship with the chaotic orbits
in the sticky zone around it (Fig.~\ref{modeorbs}f) becomes apparent.

\subsection{The spiral pattern of the model}
\label{sec:spires}
The response spiral is built by quasi-periodic orbits trapped around elliptical,
precessing backwardly with respect to the direction of rotation as energy
increases, \x periodic orbits with \ej $\geqslant -28000$. We can see a set of
these periodic orbits together with those that form the outer part of the ring
in Fig.~\ref{rsporbs}. The mechanism that reinforces the (trailing) spiral arms
is a typical ``precessing ellipses flow'' mechanism \citep{p08}. In the energy
range $-28000 <$\ej\!$< -25000$ the \x stability island occupies practically all
 the $(x,\dot{x})$ surface of section for $x>0$. The orbits that support the
spiral arms are in this case quasi-periodic orbits around \x. The orbital
dynamics are similar to those described by \citet{cg86,cg88} for open
\textit{non-barred} spirals. Corotation is placed at 10~kpc, therefore the arms
show clear bifurcating branches close the the inner 4:1 resonance. Close to
r=8~kpc the \x periodic orbits have cusps and a clear rhomboidal morphology. The
bifurcating branches are formed by the congestion of these orbits with more
circular orbits further away from the resonance.

\section{Discussion}
\label{sec:discuss}
\begin{figure*}
\begin{center}
\resizebox{140mm} {!}{\includegraphics[angle=0]{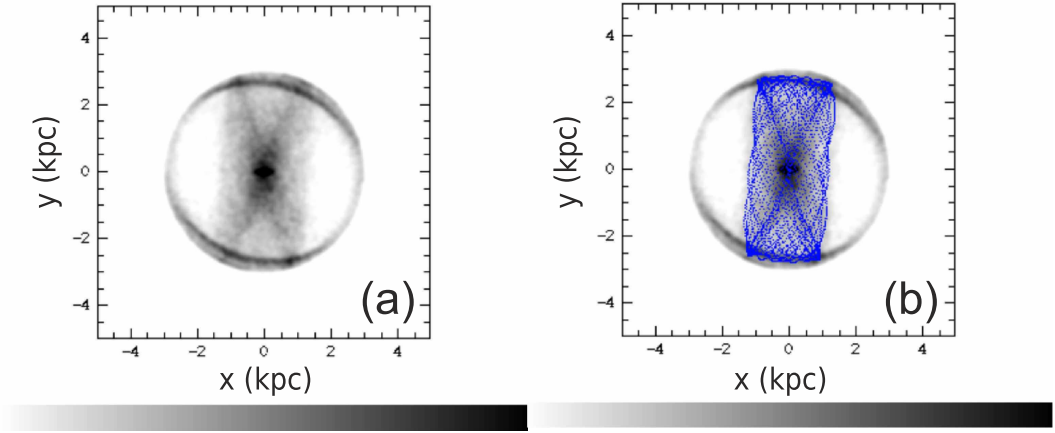}}
\end{center}
\caption{(a) The central region of Fig.~\ref{resp11}, which shows the inner
part of the ring and the X
feature in the bar of the model. (b) Two orbits from the sticky region around
the \x stability island in the surface of section for \ej $= -39000$
$(x,\dot{x})$=(0.25,-80) and $(x,\dot{x})$=(1,40) overplotted on the central
region of the model. The corresponding \x periodic orbit
is x1-like, while the chaotic orbits in this
sticky region give the orbital support to the X-shaped feature of the bar.}
\label{xorbits} 
\end{figure*}

We presented a dynamical mechanism that leads to the formation of a bar with a
characteristic X embedded in it. In our model we have simultaneously also
the formation of a ring tangent to the ends of the bar.
We summarise this result in Fig.~\ref{xorbits}.
\begin{figure*}
\begin{center}
\resizebox{160mm} {!}{\includegraphics[angle=0]{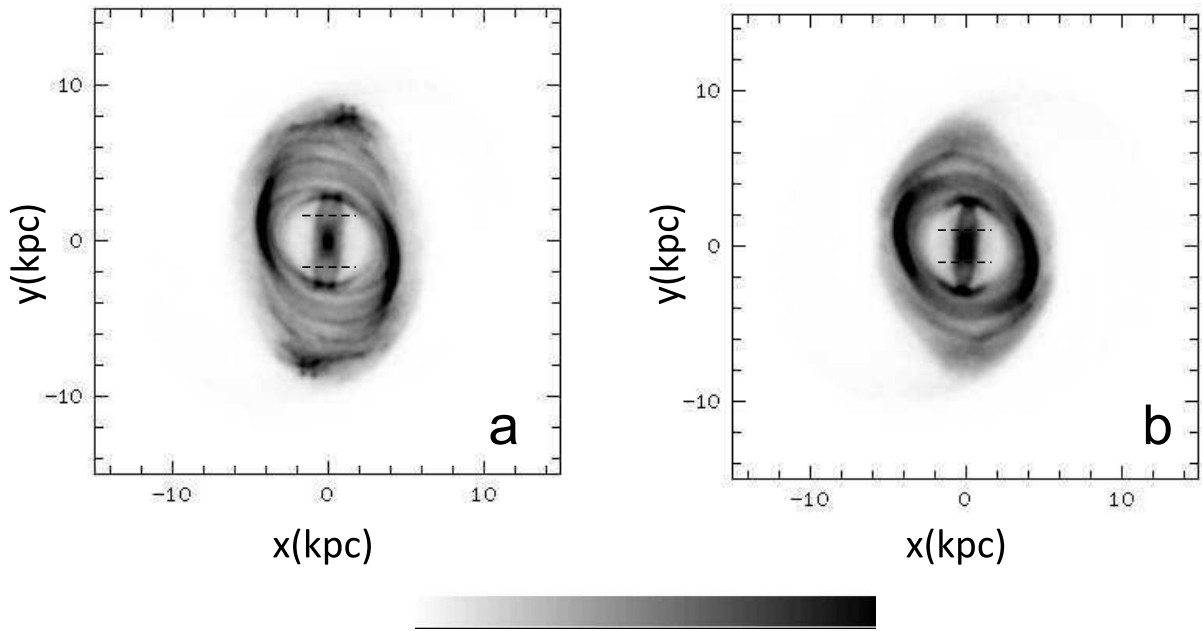}}
\end{center}
\caption {Snapshots from the response models rotating with $\varOmega_p$
=13.5~\ksk (a) and 15~\ksk (b). The dashed line segments parallel to the
x-axis indicate the height within which extends the X in the two models.} 
\label{twof} 
\end{figure*}
We have used the model with $\varOmega_p$ = 11.5~\ksk for presenting this
mechanism, because for this pattern speed the X feature extends all over the
bar. This facilitates the description of the X morphology. However, it is a
feature that we found also in the bar of other response models with $\varOmega_p
\lessapprox 15$~\ksk~\!\!, without reaching the end of the bars. We give in
Fig.~\ref{twof} two snapshots from response models with $\varOmega_p$ =
13.5~\ksk (a) and 15~\ksk (b) respectively. The maximum distances from the
x-axis within which the X extends in these two models are indicated with dashed
line segments parallel to the x-axis. The details of the X become less
discernible as $\varOmega_p$ increases. We observe that the X shrinks with
increasing $\varOmega_p$  from 11.5~\ksk (Fig.~\ref{resp11}) to 13.5~\ksk
(Fig.~\ref{twof}a) and then to 15~\ksk (Fig.~\ref{twof}b). This reflects
the amount of sticky chaotic orbits that have been populated and participate in
the building of the response bars in each case. In the models of
Fig.~\ref{twof}, at distances beyond the end of the X feature, the bar has again
a backbone of typical x1 orbits until its end.

The characteristic of the central family appears folded over a $\Delta$\ej
$\approx 2700$ range, with both x1- and x2-like flows coexisting. The curve
folds, but does not break.  A similar behaviour has been encountered also in the
slowly rotating ``Model A2" of a 3D Ferrers bar in \cite{spa02b}. The potential
we study here is not a pure bar, but a barred-spiral one, allowing both bar and
spirals appearing in the stellar response. Due to the low $\varOmega_p$,
corotation is beyond the end of the spiral arms. 

Up to the \ej of ``A'' and ``A$^{\prime}$'' our response model is populated by
orbits associated  with x1-like periodic orbits (quasi-periodic orbits from
their stability islands and chaotic orbits from the sticky zones around these
islands). For \ej larger than the \ej of ``A'' and ``A$^{\prime}$'' our model is
populated by orbits associated with the upper branch of the ``S''. For the
response model it is as ``A'' and ``A$^{\prime}$'' coincide on
the characteristic. The presence of the bistable bifurcation creates an
effective ILR, since it creates a x2-flow. The bistable bifurcation determines
the orbits that constitute the backbone of the response model. 

The ring still exists for $\varOmega_p$ = 13.5~\ksk (Fig.~\ref{twof}a) as a
pseudo-ring, or double-ring structure. A rounder inner part seems to be detached
from a more elongated x2-like outer part. In our main model (Fig.~\ref{resp11})
both parts appear joint forming a unique ring. For $\varOmega_p$ = 15~\ksk
\citep[Fig.~\ref{twof}b and][]{tp13} attached to the ends of the bar we have
only two density enhancements of ``smile'' and ``frown'' morphology. However, a
kind of loosely defined pseudo-ring structure appear even in models with
\mbox{$\varOmega_p$ = 25~\ksk\!.} The ring in our main model is formed by orbits
having energies in the middle of the \x characteristic in a (\ej\!\!,$x)$
diagram (Fig.~\ref{xar1a}) and is associated with the folding of this curve. It
is formed in the region where the central family of periodic orbits has a
bistable character. 
\begin{figure*}
\begin{center}
\resizebox{160mm} {!}{\includegraphics[angle=0]{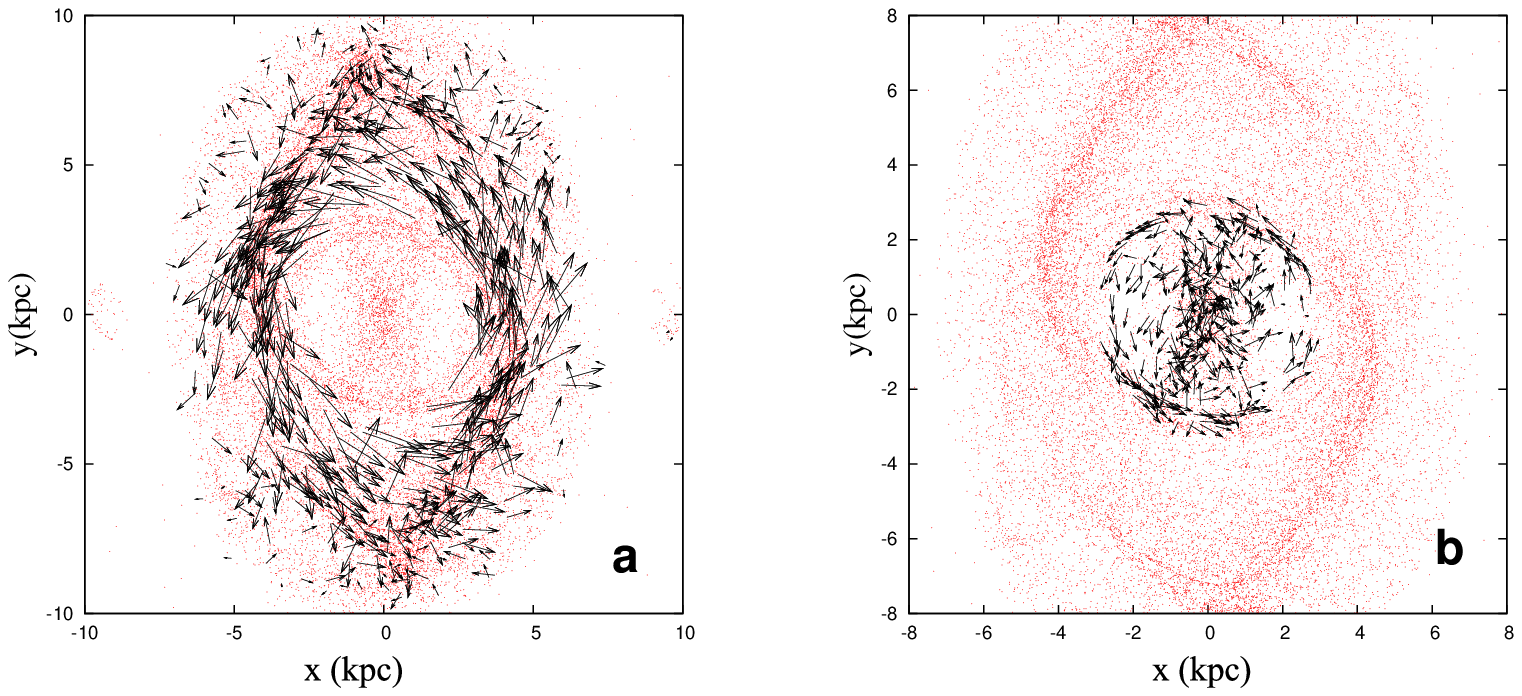}}
\end{center}
\caption{(a) The velocity field of the model outside the ring area indicates a
flow around the center of the system. (b) The velocity field of the model in the
X-shaped bar region indicates an increased dispersion of velocities.}
\label{disper} 
\end{figure*}
The ring is formed due to the introduction in the system of a x2 flow, which in
fact acts as if we had a usual inner Lindblad resonance region. By surrounding
the bar however, it should be characterised as an ``inner ring'', usually
associated with inner 4:1 or 6:1 resonances \citep[see
e.g.][]{b86,byetal94,psa03}. Because of their association with inner n:1
resonances with n$\geqq 4$, inner rings are usually orientated along the bar
\citep{b86}. In our case the ``inner'' ring can be characterised as almost
circular or rather elongated towards the minor axis of the bar, depending on the
energy level of the orbits we consider as the limit between the x2-flow orbits
and the orbits that build the spiral. 

The X in the bar is mainly a result of populating sticky orbits around the \x
stability islands in the energy range for which we have a backbone of x1-like
periodic orbits in our model. In principle this can happen independently of the
pattern speed of a model. However, in slowly rotating bars it appears
pronounced, as it occupies a larger part of them. In the case of the particular
model in this paper the ring is tangent to an ``X-shaped'' bar. Our model
describes a configuration, where ordered and chaotic motion coexist and
contribute to the formation of a unique structure. 

We note, that despite the 2D character of the present models, due to the
morphology of the orbits that build the X, we can observe a characteristic
increase in the dispersion of velocities as we cross the ring inwards
(Fig.~\ref{disper}). In Fig.~\ref{disper}a the arrows indicate a flow around the
center of the model, while in Fig.~\ref{disper}b the orientation of the orbits
looks random on the galactic plane. This is to be expected, since the motion of
the particles following X-supporting orbits is complicated. 

\citet[][pg.40]{butacat07}, \citet{lsbk11, lsabh14} give several examples of X
features in barred-spiral galaxies that are clearly far from edge-on (e.g. NGC
7020, NGC 1527, IC 5240, NGC 4429). The mechanism we propose here, i.e. the
population of sticky chaotic orbits around the stability islands of the main
family of periodic orbits, suggests an explanation of this morphology. We stress
again that the X feature is pronounced in slowly rotating models, but slow
rotation does not give the explanation of the feature. The X appears as soon as
the bar is built by sticky chaotic orbits in the chaotic zone surrounding the
stability islands of x1-like periodic orbits, as well as by quasi-periodic
orbits at the outer parts of these islands. The potential in the present
paper originates in the estimation of the gravitational field of a late-type
barred galaxy. Nevertheless, the X feature is expected to appear also in models
for early type barred galaxies, provided that the bar is populated by the kind
of sticky chaotic orbits we discuss. We note however, that there are cases of
nearly face-on galaxies combining an X feature with a ring \citep[e.g. IC
5240,][]{butacat07, lsabh14}. Such galaxies combine the morphological
features of the response model of the present study.

Recently \citet{pk14b} investigated the orbital dynamics of 3D Ferrers bars that
lead to the formation of boxy features inside the bars in their face-on
projections. Despite the fact that the bars in the two cases are approached by
different modelling techniques (2D vs. 3D dynamics), the two studies agree on
the character of the orbits that build the boxy features. In both cases we deal
with sticky chaotic orbits. Similarly with what we find here, also in the 3D
models the pool of orbits that are used to build the inner boxes are sticky
chaotic orbits around the periodic orbits that constitute the backbones of the
bars. However, the 3D modelling allows us to investigate the connection
between the X features appearing in the face-on views and the X features
appearing in boxy/peanuts edge-on profiles\citep{pk14b}.
 
\section{Conclusions}
\label{sec:concl}
The main conclusions of our study refer to the dynamics of slowly rotating
\textit{barred-spiral} models and are enumerated below:
\begin{enumerate}
\item The characteristic of the main family in our slowly rotating barred-spiral
potential includes periodic orbits that are successively (as \ej increases)
x1-like, x2-like, as well as elliptical orbits that precess if plotted at
different \ej\!. It is characterised by a folding, which we called the ``S''
(Fig.~\ref{ssos34}a). It is a case of two combined saddle-node bifurcations,
usually called ``bistability''. Along ``S'' the periodic orbits change their
orientation. Slow rotation pronounces this folding. Our response model is
populated by x1-like orbits up to the \ej of ``A'' and ``A$^{\prime}$'' in
Fig.~\ref{ssos34}a and for larger \ej by orbits associated with the upper branch
of the ``S''. For the model it is as if we jump from ``A'' to ``A$^{\prime}$''
in the characteristic.
\item The abrupt change of the orientation of the response orbits leads to the
formation of a ring in the middle of the characteristic, at the ``S'' region. 
The  ring does not extend along the bar, as is the usual orientation of inner
rings, but it is rather inclined towards the x-axis. We have effectively a x2
region beyond the end of the bar. The orientation and morphology of the ring
resembles more that of nuclear rings.
\item Bars built mainly by sticky orbits around the stability islands of the
central family of periodic orbits have a boxy structure that harbours an X
feature in it. 
In the class of barred-spiral models we consider in the present study,
the slower the pattern speed of the model, the greater is the importance of the
X-feature for the overall morphology of the bar, since it occupies a larger
fraction of it. However, the dynamical mechanism behind it is encountered in
response models within a larger range of pattern speeds, since it
depends on the amount of sticky chaotic orbits that populate the bar.
\end{enumerate}

\vspace{0.5cm} 
\noindent \textit{Acknowledgements}

We thank Prof. G.~Contopoulos for fruitful discussions and very useful comments
and the referee, Dr. Pertti Rautiainen, for his comments and suggestions, which
contributed in improving the paper. This work has been partially supported by
the Research Committee of the Academy of Athens through the project 200/823.

\label{lastpage}

\end{document}